\documentclass[twocolumn,aps,prb,showpacs,amsmath,amssymb,superscriptaddress,bibnotes,longbibliography]{revtex4-2} 
\usepackage{hyperref}
\usepackage{graphicx}
\usepackage{mathtools}
\usepackage{hyperref}
\usepackage[usenames,dvipsnames]{color}
\usepackage{diagbox}
\usepackage{pgffor}
\usepackage{tabularx}
\usepackage{tabu}
\usepackage{verbatim}
\usepackage{ifthen}
\usepackage{ulem}

\newcommand\name{}\long\def\name#1#{\romannumeral0\innername{#1}}%
\newcommand\innername[2]{%
  \expandafter\exchange\expandafter{\csname #2\endcsname}{0 #1}%
}%
\newcommand\exchange[2]{#2#1}%

\usepackage[export]{adjustbox}

\begin{document}

\title{Quantifying the role of antiferromagnetic fluctuations in the superconductivity of the doped Hubbard model}
\author{Xinyang Dong}
\affiliation{Department of Physics, University of Michigan}
\author{Emanuel Gull}
\affiliation{Department of Physics, University of Michigan}
\author{A. J. Millis}
\affiliation{Center for Computational Quantum Physics, Flatiron Institute, 162 5th Avenue, New York, NY 10010}
\affiliation{Department of Physics, Columbia University, 538 West 120th Street, New York, New York 10027}%
\begin{abstract}
    We study the contribution  of the electron-spin fluctuation coupling to the superconducting state of the two dimensional Hubbard model within dynamical cluster approximation (DCA) using a numerical exact continuous time Monte Carlo solver. By analyzing the frequency dependence of the self energy, we show that only about half of the superconductivity can be attributed to a ``pairing glue" arising from  treating spin fluctuations as a pairing boson in the standard one-loop theory.
\end{abstract}


\maketitle
\makeatletter
\let\toc@pre\relax
\let\toc@post\relax
\makeatother

\subsection{Introduction}

{Superconductivity arises from the pairing of charge $e$ electrons into charge $2e$ bosons (``Cooper pairs'') and their condensation into a coherent quantum state. In conventional superconductors such as lead, a comparison of the frequency dependence of the superconducting gap function to the frequency spectrum of the phonons (quantized lattice vibrations) \cite{Scalapino66, McMillan68} establishes that the electron-phonon interaction provides the ``pairing glue" that binds electrons into Cooper pairs.  Many ``unconventional'' superconductors are now known \cite{Steglich79,Bednorz86,Maeno94,Yoichi06} in which the pairing glue is  believed not to be provided by phonons. Substantial indirect evidence indicates that in many cases the relevant interaction is the exchange of spin fluctuations  \cite{Miyake86,Scalapino99,Maier08}, 
but direct evidence has been lacking and many other mechanisms have been proposed \cite{Castellani96,Varma97,Capone04,Anderson07,Stanescu08,Saito15}.

}

The theoretical study of the 
unconventional superconductivity that is believed to arise from strong electron-electron interactions requires a model that captures the essentials of the correlated electron physics, and can be studied non-perturbatively. The Hubbard model \cite{LeBlanc15, Zheng17} has been proposed as the minimal theoretical model of quantum materials such as the copper-oxide based high-T$_c$ superconductors \cite{Anderson87}. 
This model describes electrons hopping among sites of a lattice (here we consider the two dimensional square lattice case with nearest-neighbor hopping of amplitude $t$) and subject to a site-local repulsive interaction $U$.

To have non-perturbative access to both the static phase diagram and dynamical properties, we use the dynamical cluster approximation (DCA) \cite{Maier05} method. In DCA, the electron propagator and spin fluctuation spectrum are computed within the same formalism and at the same level of approximation, enabling a quantitative analysis of the electron-spin fluctuation interaction. The resulting solution  \cite{Gull10,Gull13} produces a good qualitative description of the physics of the high-T$_c$ copper oxide superconductors, including a high-doping Fermi liquid regime, a Mott insulator, a low doping pseudogap and an intermediate-doping dome of d-wave superconductivity. The extent to which a stripe magnetic phase preempts the superconducting phase found in the DCA is currently under debate \cite{Zheng17,Qin20}, but we emphasize that the superconductivity found in DCA is well defined and locally stable, with properties that we study in this paper. 

We quantify the strength of the electron-spin fluctuation coupling in the model by analysing the frequency dependence of the computationally determined electron self energy, superconducting gap function and spin fluctuation spectrum. 
Our analysis shows that at intermediate interaction and slightly overdoped regime,
about half of the superconductivity is attributable to spin fluctuations in the  one-loop spin fluctuations, with the other half coming from higher energy processes.

\subsection{Results}
We investigated several different dopings and interaction strengths. We present here results obtained for doping $x \sim 0.10$ (carrier concentration $n=1-x$ per site) and temperatures as low as $T=t/50$. For this carrier concentration at $U=6t$, the normal state is a momentum-space differentiated Fermi liquid outside the pseudogap regime, corresponding to the overdoped side of the cuprates. The superconducting state, which we explicitly construct, appears below a transition temperature $T_c\approx t/40$.
The choice of parameters is influenced by the following considerations: for higher $U$, calculations become more difficult \cite{Gull08}, while for lower $U$ they are less relevant for strong correlation superconductivity. Higher dopings reduce $T_c$, whereas lower dopings enhance the effects of the nearby pseudogap and the effects of the AFM state around half filling, making  one-loop spin fluctuation theory less likely to succeed. We will comment briefly on the results for different dopings in the conclusions.

We calculate the normal (N) and anomalous (A) components of the electron self energy. Using  recent algorithmic developments \cite{Chen15} we also calculated the impurity model spin susceptibilities $\chi$  in both normal and superconducting states. 

\begin{figure}[tbh]
\centering
\includegraphics[width=0.95\columnwidth]{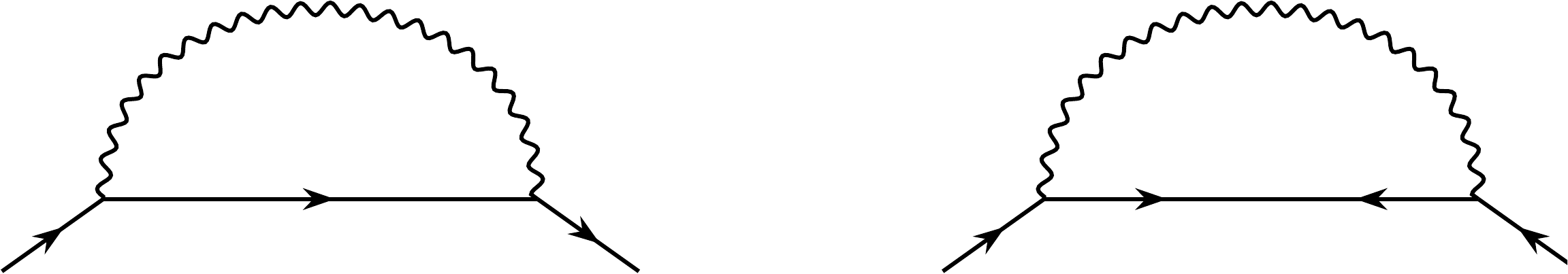}
\caption{
    Spin-fluctuation diagrams for normal and anomalous self energy.
    Solid lines: Normal or anomalous Green's function; Wavy lines: Magnetic susceptibility.
    }\label{fig:diagram}
\end{figure}

Spin fluctuation theories yield the spin fluctuation (SF) contribution to the normal (N) and anomalous (A) self energies in terms of the spin susceptibility and normal and anomalous components of the Green function $G_{K}^{N/A}$ as \cite{Scalapino66,Anderson73,Miyake86, Maier08} (see Fig.~\ref{fig:diagram})
\begin{equation}
    \Sigma_{K}^{SF;N/A}(\omega)=g^2 \frac{1}{\beta N}\sum_{Q,\Omega}\chi_Q(\Omega)G_{K-Q}^{N/A}(\omega-\Omega).
    \label{eq:sigma}
\end{equation}
We assess the relevance of spin fluctuations by using our calculated $G$ and $\chi$, along with an estimated coupling constant $g$ to compare $\Sigma^{SF}$ (Eq.~\ref{eq:sigma}) to our numerically calculated self energies.
\begin{figure}[htbp]
\centering
\includegraphics[width=0.95\columnwidth]{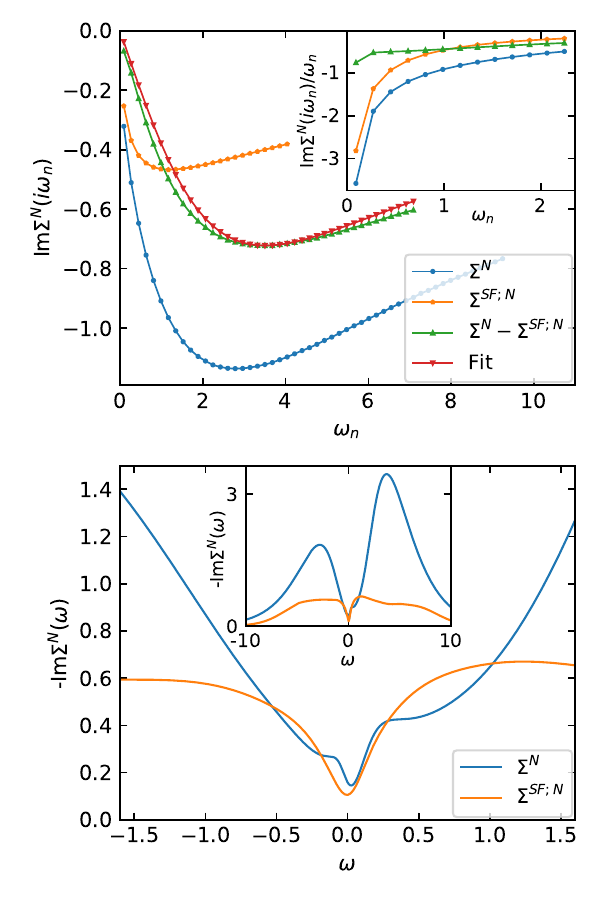}
\caption{Upper panel: Imaginary part of the normal component of the Matsubara self energy for the antinode $(0,\pi)$ at $U=6$, $\beta=35$, and $\mu=-1.0$ compared to spin fluctuation self energy computed with $g^2=3.8$. 
Inset: $\text{Im}\Sigma^N_K(i\omega_n)/\omega_n$.
Lower panel: Negative of the analytically continued real-axis the antinode $(0,\pi)$ self-energy and spin-fluctuation contribution computed with $g^2=3.8$. 
Inset: Self energy over a wide frequency range.
\label{fig:ImSigmaN} 
}
\end{figure}

The method to estimate $g^2$ is explained in detail in the method section.
In general, we partition the imaginary part of the real frequency self energy into a low frequency part that we suppose arises mainly from spin fluctuations and a higher frequency part that represents all of the other processes contributing to the imaginary part of the self energy: $\text{Im}\Sigma^{N}_K(\omega)=\text{Im}\Sigma^{SF;N}_K(\omega)+\text{Im}\Sigma^{\text{high;N}}_K(\omega)$. We take $\Sigma^{SF;N}$ to have the functional form of Eq.~\ref{eq:sigma} and determine $g^2$ by requiring consistency with our numerically computed self energies. We have computed the self energies in all momentum tiles but focus here on the self energies corresponding to the tiles centered on the antinode points $(\pi,0)/(0,\pi)$, where the superconducting gap is maximal and the normal component of the  self energy is largest. We consider consistency both directly on the Matsubara axis (avoiding the ambiguities associated with analytical continuation) and on the real axis. 
For the imaginary axis analysis we note that the quantity $Z^N_K=\frac{\partial [\text{Re}\Sigma^N_K(\omega)]}{\partial \omega}$ related to the normal state mass enhancement may be estimated from Matsubara axis results as  $Z^N_K = \frac{\text{Im}(\Sigma^N_K(i\omega_1) - \Sigma^N_K(i\omega_0))}{\omega_1 - \omega_0}$ (see inset of Fig.~\ref{fig:ImSigmaN}) and cannot be larger than the contribution from the spin fluctuation sector. 

The upper panel of  Fig.~\ref{fig:ImSigmaN} shows the Matsubara analysis of the normal component of the antinode self energy and the lower panel shows the real axis fits. Both cases are consistent with a value of $g^2=3.8$ implying that about $2/3$ of $Z^N$ comes from the electron-spin fluctuation interaction. 

\begin{figure}[tbh]
    \centering
    \includegraphics[width=0.95\columnwidth]{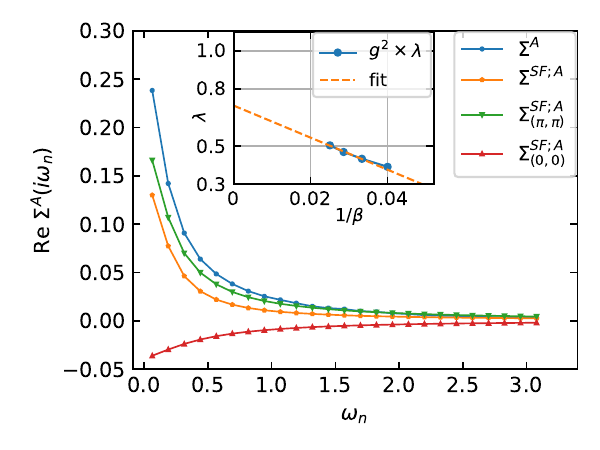}
    \caption{
    Total measured anomalous self-energy $\Sigma^A_K$ and estimated spin-fluctuation contribution $\Sigma^{SF; A}_K$ of $K=(0, \pi)$ at $U = 6$, $\beta = 50$ and $\mu = -1.0$ ($n = 0.90$). Also shown are the individual contributions to $\Sigma^{SF; A}$ from  transferred momenta $Q=(\pi,\pi)$ and $Q=(0, 0)$.
    Inset: Leading eigenvalues computed from the linearized self energy equation Eq.~\ref{eq:sigmaA_linear}. The value of $g^2$ is chosen to be 3.8 for all temperatures. Dotted line: Linear fit to $\beta=30, 35, 40$.
    \label{fig:ReSigmaA}
    }
\end{figure}

With the spin fluctuation spectrum and the electron-spin fluctuation coupling constant in hand, we next determine the extent to which superconductivity arises from spin fluctuations by solving the anomalous component of Eq.~\ref{eq:sigma} and comparing the result to the numerically exact CTQMC solution which gives $d_{x^2-y^2}$ symmetry  superconductivity. We begin with the equation for the transition temperature $T_c$, obtained by linearizing Eq.~\ref{eq:sigma} in the anomalous component of the self energy. The resulting equation is a linear eigenvalue equation for eigenvector $\Sigma^A(i\omega_n)$; the largest eigenvalue $\lambda$ increases as temperature decreases, and $T_c$ is the temperature at which the leading eigenvalue equals unity [see Eq.~\ref{eq:sigmaA_linear}]. A $d_{x^2-y^2}$ symmetry gap yields a non-negative eigenvalue. Using our estimated $g^2=3.8$, we find that at temperature $T=t/40$ the leading eigenvalue $\lambda$ is about $0.5$ [see inset of Fig.~\ref{fig:ReSigmaA}], so that increasing the net pairing strength by a factor of about two would be needed to bring the leading eigenvalue up to $1$ (in fact a larger increase would be required because the coupling constant of the normal state self energy means the transition temperature does not vary linearly with the coupling).

Figure.~\ref{fig:ReSigmaA} compares the QMC anomalous self energy to the spin fluctuation self energy $\Sigma^{SF; A}_K$ at $K=(0, \pi)$. We note that the spin fluctuation interaction has two components, one from fluctuations near the antiferromagnetic  wavevector $(\pi,\pi)$ and one from fluctuations at small momenta near $Q=(0,0)$. The small momentum fluctuations make a negative contribution to $\Sigma^A_K$. At the lowest Matsubara frequency the $\Sigma^{SF; A}_K$ produced by the spin fluctuation theory is approximately half of the QMC self energy, again indicating that spin fluctuation theory \cite{Scalapino66} alone cannot account for the superconductivity.

\begin{figure}[tbh]
    \centering
    \includegraphics[width=0.95\columnwidth]{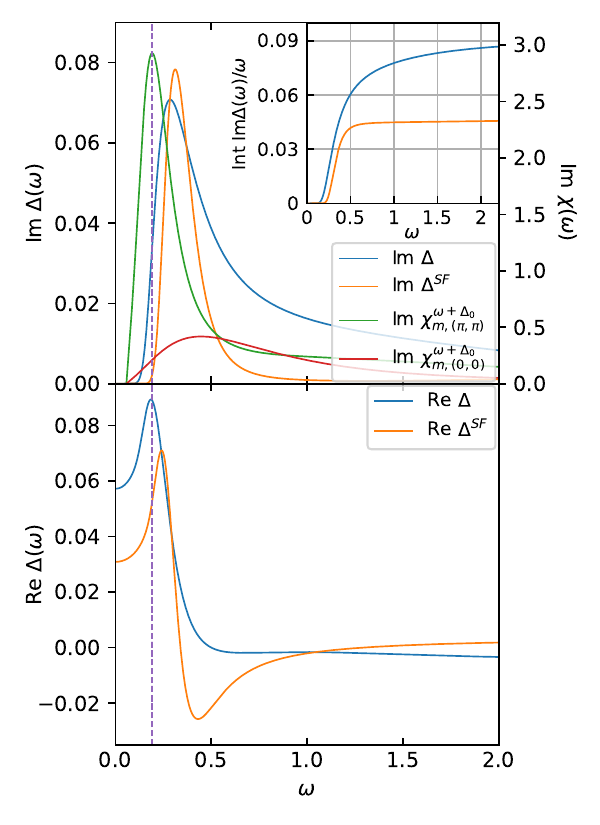}
    \caption{
    Comparison of the true gap function, gap function from spin fluctuation, the AFM susceptibility $\text{Im}\chi^{\omega + \Delta_0}_{m, (\pi, \pi)}$ and the FM susceptibility $\text{Im}\chi^{\omega + \Delta_0}_{m, (0, 0)}$ shifted by the $\Delta_{0} = \text{Re} \Delta(\omega=0) = 0.057$ at $U = 6$, $\beta = 50$ and $\mu = -1.0$ ($n = 0.90$). Upper panel: Imaginary part. Inset: Integral of $\text{Im} \Delta(\omega)/\omega$ starting from $\omega = 0$. Lower panel: Real part. 
    \label{fig:gaps}
    }
\end{figure}

We now examine in Fig.~\ref{fig:gaps}  the frequency dependence of the gap function $\Delta(\omega)$, a complex function of real frequency  defined in terms of the normal and anomalous self energies at $K = (0, \pi)$ as \cite{Poilblanc02,Gull14}
\begin{align}
    \Delta(\omega)=\Sigma^A_K(\omega)\Big/\left[1-\frac{\Sigma^N_K(\omega)-\Sigma^N_K(-\omega)}{2\omega}\right] \, .
\end{align}
Following Ref.~\cite{Scalapino66} we compare the frequency dependence of the spin fluctuation spectrum, the imaginary part of the DCA-computed gap function, and the estimated gap function computed by solving  Eq.~\ref{eq:sigma} using the CT-QMC-computed $\Sigma^A$ and $\chi$. The real frequency quantities are obtained from maximum entropy analytical continuation of imaginary frequency data obtained at $T=t/50$, well below the superconducting transition temperature. As noted in Ref.~\cite{Scalapino66} the presence of a gap in the electron Greens function means that a peak in $\chi$ at a frequency $\omega_{peak}$ implies a peak in $\Delta$ at $\omega+\omega_{peak}$ so we shift $\chi$ by the zero frequency gap function in the comparison. 

We emphasize that the uncertainties in the analytical continuation are not small; while areas are reliably estimated, peak heights and widths are subject to some uncertainty. We see from Fig.~\ref{fig:gaps} that while the peaks in the gap function and shifted $\chi$ roughly coincide, the spin fluctuation contribution to the imaginary part of the gap function is concentrated at low frequencies, decaying much more rapidly than the DCA-computed gap function, further demonstrating the importance of a high-frequency non-spin-fluctuation contribution to the electron self energy.

\subsection{Discussion}
Spin fluctuation theories, in which the spin fluctuations (as parametrized by the susceptibility) are treated as a pairing boson within the one loop approximation, are widely considered to be promising candidates for theories of superconductivity. Here we have performed a quantitative study, in a well defined, numerically controlled theory, of the extent to which this is actually the case. The theory produces a superconducting  state and a spin fluctuation spectrum, which (taking  advantage of recent developments \cite{Chen15}) we can obtain numerically exactly.   Access to the spin fluctuation spectrum enables us to compare the spin fluctuation theory calculation of the normal state self energies to numerically exact results for the same quantities, thereby allowing an estimate of the electron-spin fluctuation coupling constant. Knowledge of the coupling constant then enables a quantitative analysis of the contribution of spin fluctuations to the superconducting transition temperature and to the magnitude and form of the superconducting gap function. In qualitative consistency with previous results \cite{Maier08} we find that low-frequency spin fluctuations contribute to the superconductivity, but we find that quantitatively only about half of the pairing can be attributed to these fluctuations. 
The other half of the pairing therefore arises from higher frequency fluctuations, whose nature and precise physical origin remains to be determined.

We have similarly examined other doping values and interaction strengths, including $U=6,~\mu=-0.9~(x \sim 0.088)$, $U=6,~\mu=-1.1~(x \sim 0.12)$, and $U=5.5,~\mu=-0.6~(x \sim 0.066)$.  For these parameters, our  analysis works. We found that as doping is decreased, spin fluctuation theory rapidly becomes a much less satisfactory description of the normal state, with the spin fluctuation contribution to $\Sigma^N$ apparently decreasing, whereas the transition temperature weakly increases. As the doping is increased, the spin fluctuation contribution to the normal state self energy and gap function becomes larger, but the transition temperature rapidly decreases. These two results confirm that spin fluctuations do not fully account for the superconductivity exhbited by the model.

The theoretical model used in this work is the 8-site cluster dynamical mean field approximation, in the `DCA' implementation. The cluster size is chosen based on previous literature to capture the pairing and magnetic fluctuations at reasonable computational expense.  Cluster dynamical mean field theory does not adequately capture for example the stripe physics \cite{Zheng17,Huang18,Qin20,Wietek21,Mai21} that may preempt superconductivity in some parameter ranges, and the cluster sizes available, while large enough to provide results that compare well to experiment and more exact calculations, cannot capture many of the interesting specifics of superconducting phenomenology. However, it is important to emphasize that the method provides a single internally consistent computational scheme that produces a well defined locally stable superconducting phase whose properties can be studied, and that provides, at the same level of approximation, normal and anomalous self energies and spin fluctuation spectra, enabling a theoretically meaningful comparison.  

Our finding that spin fluctuations, as parametrized by the spin-spin correlation function $\chi$, and coupled to electrons via the standard one-loop approximation, are not the dominant form of superconductivity suggests more generally that spin fluctuation theories of this type may miss important aspects of correlated electron superconductivity.
Our finding also suggests that if the nature of the higher frequency contributions to the pairing could be elucidated, tuning these degrees of freedom might be an effective strategy for raising the transition temperature.

\subsection{Methods}
\subsubsection{Hubbard model, self energy, and magnetic susceptibility}
We study the two dimensional single band Hubbard model in both the normal and the superconducting state
\begin{align}
    H = \sum_{k\sigma} (\epsilon_k - \mu) c_{k\sigma}^\dagger c_{k\sigma} + U \sum_i n_{i\uparrow} n_{i\downarrow} \, ,
\end{align}
with $\mu$ the chemical potential, $\epsilon_k=-2t(\cos k_x + \cos k_y)$ the dispersion with nearest neighbor hopping $t$. $U$ is the strength of the interaction, $i$ labels a lattice site, $k$ labels the momentum, and $n$ is the density operator.
\\
We measure the Green's function matrix $\underline{G}(k, i\omega_n)$ in the impurity solver, and
the self energy can be computed from the Dyson equation
\begin{align}
    \underline{\Sigma}(k, i\omega_n) =  \underline{{G}}^{-1}_{0}(k, i\omega_n) - \underline{{G}}^{-1}(k,i\omega_n) \, ,
\end{align}
with
\begin{align}
    &\underline{G}(k, \tau) =
    -\langle \mathcal{T}
    \begin{pmatrix}
    c_{k\uparrow}(\tau)c^\dagger_{k\uparrow}(0) & c_{k\uparrow}(\tau)c_{-k\downarrow}(0)\\[0.3em]
    c^\dagger_{-k\downarrow}(\tau)c^\dagger_{k\uparrow}(0) & c^\dagger_{-k\downarrow}(\tau)c_{-k\downarrow}(0)
    \end{pmatrix}
    \rangle \, ,
    \\
    &\underline{G}(k, i\omega_n) 
    = \int_0^\beta d\tau e^{i\omega_n \tau} \underline{{G}}(k, \tau) \nonumber \\
    &\phantom{\underline{G}(k, i\omega_n)}=\begin{pmatrix} 
    G^N_{k\uparrow}(i\omega_n) &G^A_{k\uparrow}(i\omega_n) \\[0.3em]
    G^{A*}_{k\uparrow}(i\omega_n) &-G^N_{-k\downarrow}(-i\omega_n)
    \end{pmatrix} \, ,
    \\
    &\underline{G}_0^{-1}(k, i\omega_n) =
    \begin{pmatrix} 
    i\omega_n - \epsilon_k + \mu &0 \\[0.3em]
    0 & i\omega_n + \epsilon_k - \mu
    \end{pmatrix} \, ,
    \\
    &\underline{\Sigma}(k, i\omega_n) = 
    \begin{pmatrix} 
    \Sigma^N_{k\uparrow}(i\omega_n) &\Sigma^A_{k\uparrow}(i\omega_n) \\[0.3em]
    \Sigma^{A*}_{k\uparrow}(i\omega_n) &-\Sigma^N_{-k\downarrow}(-i\omega_n)
    \end{pmatrix} \, .
\end{align}
The SU(2) symmetry of the system gives $G^N_{\uparrow} = G^N_{\downarrow}$.
\\
The magnetic susceptibility is defined with the correlator of the magnetization in $z$ direction $\hat{S}_z = n_{\uparrow} - n_{\downarrow}$
\begin{align}
    &\chi_m(q, \tau) 
    = \langle \mathcal{T} \hat{S}_z(q, \tau) \hat{S}_z(-q, 0) \rangle -  \langle \hat{S}_z(q) \rangle^2 \, , \\
    &\chi_m(q, i\Omega_n) 
    = \int_0^\beta d\tau e^{i \Omega_n \tau} \chi_m(q, \tau) \, .
\end{align}
We measure $\chi_m(q, \tau)$ on the Chebyshev-Gauss-Lobatto collocation points and compute $\chi_m(q, i\Omega_n)$ via spectral transform \cite{Shen11,Gull18}.

\subsubsection{Numerical method}
We use the dynamical cluster approximation (DCA) to compute the single particle Green's function and the susceptibility.
The DCA \cite{Maier05} proceeds by tiling the Brillouin zone into $N$ equal-area non-overlapping tiles $a$ centered at momentum points $K_a$ and approximating the electron self energy as $\Sigma_k(i\omega_n)\rightarrow \Sigma_{K_a}(i\omega_n)$ for $k$ in tile $a$, so that the momentum dependence is approximated as a sum of piecewise constant functions and the full frequency dependence is retained. The $\Sigma_{K_a}(i\omega_n)$ are obtained from the solution of a $N$-site quantum impurity model with the same interaction $U$ as in the original model and single particle parameters obtained by a self-consistency condition. We have chosen $N=8$ which provides sufficient momentum resolution while allowing for calculation of the detailed dynamical information needed here.  

The impurity model is solved with the continuous-time quantum Monte Carlo methods \cite{Gull08,Gull11}.

\subsubsection{Coupling constant}
We compute the one-loop spin fluctuations in Matsubara frequency space via
\begin{align}
        \Sigma_{K}^{SF;N/A}(i\omega_n)= \frac{g^2}{\beta N}\sum_{Q,i\Omega_n}\chi_Q(i\Omega_n)G_{K-Q}^{N/A}(i\omega_n-i\Omega_n).
    \label{eq:sigma_sf}
\end{align}
To estimate the coupling constant $g^2$, we partition the exact normal self energy from DCA into a a low frequency part that is supposed to arise mainly from spin fluctuations, and a higher frequency part that represents contributions from all other processes
\begin{align}
    \text{Im}\Sigma^{N}_K(i\omega_n)=\text{Im}\Sigma^{SF;N}_K(i\omega_n)+\text{Im}\Sigma^{\text{high;N}}_K(i\omega_n) \, , \label{eq:sigma_high}
\end{align}
where the high frequency process is fitted by a minimal two-parameter equation 
\begin{align}
    \text{Im}\Sigma_K^{\text{fit};N}(i\omega_n) = -\frac{A}{\pi} \frac{\omega_n}{\omega_n^2 + x_0^2} \, ,
    \label{eq:sigma_high_fit}
\end{align}
with $A$ and $x_0$ being two fitting parameters.
The other relation we impose in the fitting procedure is that the quasi-particle weight $\left[1-\frac{\partial [\text{Re}\Sigma^N_K(\omega)]}{\partial \omega} \right]^{-1}$ given by the exact self energy and the approximated self energy from the spin fluctuation plus the high frequency fitting are approximately the same.
\begin{subequations}
\begin{align}
    &Z^N_K = Z^{\text{fit}; N}_K + Z^{SF;N}_K \, , 
    \\
    &Z^N_K = \frac{\text{Im}(\Sigma^N_K(i\omega_1) - \Sigma^N_K(i\omega_0))}{\omega_1 - \omega_0} \, .
\end{align}
\label{eq:Z}
\end{subequations}
The fitting procedure is as follows:
\begin{itemize}
    \item For a given $g^2$, compute $\Sigma_K^{SF;N}(i\omega_n)$ as in Eq.~\ref{eq:sigma_sf}.
    \item Compute Im$\Sigma_K^{\text{high};N}(i\omega_n)$ as in Eq.~\ref{eq:sigma_high}.
    \item Fit Im$\Sigma_K^{\text{fit};N}(i\omega_n)$ to Im$\Sigma_K^{\text{high};N}(i\omega_n)$ by computing the two fitting parameters $A$ and $x_0$ from the maximum of Im$\Sigma_K^{\text{high};N}(i\omega_n)$.
    \item Compute $\bar{g}^2$ from the requirement of Eq.~\ref{eq:Z}.
\end{itemize}
The value of $g^2$ is decided by requiring $g^2 = \bar{g}^2$ in the above procedure, under the constraint $A>0$, $-\text{Im}\Sigma^{SF;N}_K(i\omega_n) \leq -\text{Im}\Sigma^{N}_K(i\omega_n), \, \forall n$, $-Z^{SF;N}_K \leq -Z^N_K$.

\subsubsection{Linearized self energy equation}
From the matrix form of the Dyson equation, the linearized anomalous Green's function can be computed as
\begin{align}
    G^A_K(i\omega_n) = \frac{\text{Im} G^N_K(i\omega_n)}{\omega_n - \text{Im}\Sigma^N_K(i\omega_n)} \Sigma^A_K(i\omega_n) \, .
\end{align}
In an eight-site DCA simulation with d-wave superconductivity, the anomalous Green's function and self energy will only be non-zero at $K = (0, \pi)$ and $(\pi, 0)$ and $G^A_{(0, \pi)}(i\omega_n) = -G^A_{(\pi, 0)}(i\omega_n)$, $G^N_{(0, \pi)}(i\omega_n) = G^N_{(\pi, 0)}(i\omega_n)$.
The one-loop spin fluctuations Eq.~\ref{eq:sigma_sf} can then be rewritten as
\begin{align}
    \Sigma^A_{(0, \pi)}(i\omega_n)& \nonumber \\
    = \frac{g^2}{\beta N} \sum_{\omega_n'}
    &\left[
    \chi_{(0, 0)}(i\omega_n - i\omega_n') - \chi_{(\pi, \pi)}(i\omega_n - i\omega_n')
    \right] \nonumber \\[-0.8em]
    &\times \frac{\text{Im} G^N_{(0, \pi)}(i\omega_n')}{\omega_n' - \text{Im}\Sigma^N_{(0, \pi)}(i\omega_n')} \Sigma^A_{(0, \pi)}(i\omega_n') \nonumber \\
    = \sum_{\omega_n'} & F(i\omega_n, i\omega_n')\Sigma^A_{(0, \pi)}(i\omega_n') \, ,
    \label{eq:sigmaA_linear}
\end{align}
where $F(i\omega_n, i\omega_n')$ is a matrix in $\omega_n$ and $\omega_n'$. The leading eigenvalue $\lambda$ of this matrix should cross one at T$_c$, if spin-fluctuations of this form cause superconductivity, and otherwise denotes the fraction of superconductivity given by one-loop spin fluctuations.

\acknowledgements{
EG and XD are supported by NSF DMR 2001465. The Flatiron Institute is a division of the Simons Foundation.}

\bibliographystyle{apsrev4-2}
\bibliography{refs}
\end{document}